\documentclass[a4paper,showpacs,twocolumn,amsmath,amssymb,floatfix,prb,preprintnumbers,footinbib]{revtex4}
\usepackage{graphicx}
\usepackage{amsmath,amsfonts}
\usepackage{color}

\newcommand{\e}{\varepsilon}
\newcommand{\m}[1]{\mathrm{#1}}
\newcommand{\ket}[1]{\left|#1\right\rangle}
\newcommand{\bra}[1]{\langle #1|}

\begin{document}

\title{Adiabatic pumping in a double-dot Cooper-pair beam splitter}

\author{Bastian Hiltscher$^1$, Michele Governale$^2$, Janine Splettstoesser$^{3}$, and J\"urgen K\"onig$^1$} 
\affiliation{$^1$Theoretische Physik, Universit\"at Duisburg-Essen and CeNIDE, D-47048 Duisburg, Germany\\
$^2$School of Chemical and Physical Sciences  and MacDiarmid Institute for Advanced Materials and Nanotechnology, Victoria University of Wellington, P.O. Box 600, Wellington 6140, New Zealand\\
$^{3}$Institut f\"ur Theorie der Statistischen Physik, RWTH Aachen University, D-52056 Aachen, and JARA-Fundamentals of Future Information Technology, Germany}

\date{\today}
\begin{abstract}
We study adiabatic pumping through a double quantum dot coupled to normal and superconducting leads.  
For this purpose a perturbation expansion in the tunnel coupling between the dots and the normal leads is performed and processes underlying the pumping current are discussed.  
Features of crossed Andreev reflection are investigated in the average pumped charge and related to local Andreev reflection in a single quantum dot.
In order to distinguish Cooper-pair splitting from quasi-particle pumping, we compare the properties of Cooper-pair pumping with single-electron pumping in a
system with only normal leads. The dependence on the average dot level and the coupling asymmetry turn out to be the main distinguishing features. This is contrasted with the linear conductance for which it is more difficult to distinguish single-particle from Cooper-pair transport. 
\end{abstract}


\maketitle
\section{Introduction}
Charge transport through interfaces between superconductors and normal conductors takes place by different processes. Above the energy gap of the superconductor's density of states, mainly single electrons are transferred, while subgap transport is sustained by Andreev reflection (AR).~\cite{andreevreflection} In an AR process, an electron in the normal conductor that impinges on the interface is retroreflected as a hole while a Cooper-pair is transferred into the superconductor. In junctions with more than one normal conductor also \textit{crossed Andreev reflection} (CAR) may occur, that is, the two electrons  forming the Cooper-pair stem from different normal conductors (or tunnel into different normal contacts in the opposite transport process). This nonlocal transport mechanism has been extensively studied both theoretically~\cite{CARtheory} and experimentally.~\cite{CARexperiments} 

\begin{figure}
\vspace{-9cm}
\includegraphics[angle=270,width=1.1\textwidth]{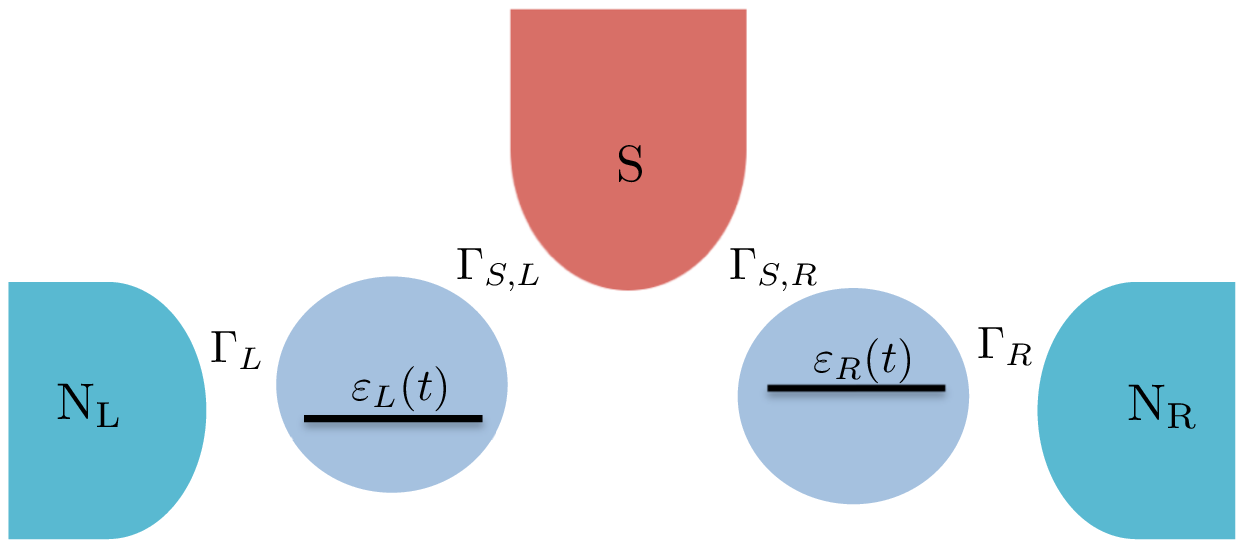}
\caption{
\label{setup}
(Color online) NDSDN setup: two quantum dots coupled to the same superconductor and each dot coupled to a normal conductor.}
\end{figure}

In recent years advancements in nanofabrication have made it possible to contact quantum dots (QDs) with superconducting leads.~\cite{SCQDdevices,hofstetter09,herrmann10} Such QD-superconductor devices are of great relevance, because they enable the investigation of the interplay between superconducting correlations and Coulomb repulsion in nonequilibrium situations. Andreev reflection~\cite{QDARtheory,FDStheory,pala07,futterer10} as well as crossed Andreev reflection~\cite{futterer09,QDCARtheory,eldridge10} through quantum dots have been the focus of many theoretical works. Recently, CAR through QDs has also been observed in experiment.~\cite{hofstetter09,herrmann10} The setup consists of a superconducting lead tunnel coupled to two parallel quantum dots realized in an InAs semiconducting nanowire~\cite{hofstetter09} and a carbon nanotube,~\cite{herrmann10} respectively.  Each of the two quantum dots is additionally coupled to separate normal reservoirs (see Fig.~\ref{setup}). As a result the dependence of the current in one arm of the beam splitter on the parameters of the other arm indicates the occurrence of CAR.

In the examples mentioned above, a bias voltage is applied to generate dc transport. In this paper we consider a different transport mechanism: \textit{adiabatic pumping}. The principle of pumping is to transport electrons in the absence of a bias voltage by varying certain system parameters periodically in time. Pumping is therefore a mechanism converting an ac into a dc signal, which has been experimentally realized in different
systems.~\cite{pumpexp}  In the adiabatic regime the pumping period is large compared to other characteristic time scales of the system. 
It was shown~\cite{splettstoesser06,reckermann10} that adiabatic pumping reveals features which are not visible in stationary transport. Here, the main motivation of our work is to use adiabatic pumping in order to investigate features of CAR.
We therefore consider a system consisting of  two quantum dots coupled both to normal and superconducting leads as shown in Fig.~\ref{setup}. In the experiments performed so far,  the CAR and AR signals coexist. This happens even though strong Coulomb interaction within each dot tends to suppress AR, therefore enhancing the visibility of CAR. Adiabatic pumping requires two out-of-phase  time-dependent parameters in order to obtain a finite dc current.  Choosing gates applied to the two dots, belonging to two different arms of the beam splitter, as pumping parameters, only transport mechanisms relying on nonlocal correlations between the two arms contribute to the pumped charge. Therefore, such a type of pumping cycle has the advantage with respect to biased transport that it singles out CAR, while local effects do not yield any finite dc current.\\
Theoretically, the dynamical scattering approach provides a general framework for pumping as long as the Coulomb interaction is weak.~\cite{buettiker94} In noninteracting systems, the influence of the superconducting proximity effect on pumping was studied in Refs.~\onlinecite{wang01}.
However, Coulomb interaction cannot be neglected in the setup considered here. In recent years much effort has been put on the treatment of  pumping through strongly interacting systems such as quantum dots.~\cite{pumpingQDtheory} While pumping through a single quantum dot with a superconducting lead was studied in the limit of zero temperature and infinitely strong Coulomb interaction,~\cite{splett07} in this paper, we are interested in systems in which Coulomb interaction can be arbitrary and where coupling to the leads is weak. To this purpose we use an adiabatic extension of a generalized master equation approach.~\cite{splettstoesser06,reckermann10,adexpgm,riwar10}  In the stationary limit the generalized master equation approach~\cite{konig96} has been applied to hybrid quantum dot systems before.~\cite{pala07,futterer09,eldridge10}

The motivation of this paper is to identify and understand CAR in adiabatic pumping. To this purpose we investigate two quantum dots, with infinite intra-dot Coulomb repulsion,  tunnel coupled to the same superconductor and  each of them to a normal conductor (NDSDN) (see Fig.~\ref{setup}). Pumping is realized by applying time-dependent potentials, namely one to each of the quantum dots, via gates with a phase-difference in the driving. This gives us the possibility to identify unique features of crossed Andreev reflection in adiabatic pumping which rely on the \textit{nonlocality} of the effect and can - as we show by a comparison with a setup with the superconductor replaced by a normal lead (NDNDN)  - not be reproduced by other parasitic nonlocal effects mediated by quasiparticles.\\
However, the complexity of this setup makes it difficult to obtain compact analytic formulae. Therefore, we additionally consider a quantum dot with Zeeman-split levels, tunnel coupled to a ferromagnetic and a superconducting lead (FDS). In hybrid systems containing ferromagnets, superconductors, and quantum dots the influence of spin asymmetry on Andreev reflections has been investigated before.~\cite{FDStheory,futterer09,futterer10,hofstetter10} In the present work our motivation of considering the FDS setup is to get a better understanding of the transport processes in the NDSDN system because we can relate the CAR in the NDSDN setup to AR in the FDS setup. The Zeeman splitting and the polarization in the FDS setup corresponds to a difference of the two dot levels  and an asymmetry of the coupling to the two normal conductors of the NDSDN system, respectively. From a theoretical point of view the main difference between the two setups is the existence of 
triplet states in the NDSDN system. Experimentally, although hybrid systems containing ferromagnets and superconductors are realizable,~\cite{hofstetter10} the time dependence of the transport channels through the dot are easier to control in the NDSDN setup.

The paper is structured as follows. In Sec.~\ref{model} we present the models of the considered setups.  The technique used to compute the pumping current into the superconductor is described in Sec.~\ref {method}. The results divide in three different parts. In Secs.~\ref{secnds} and \ref{secndsdn} the results for local Andreev reflections and crossed Andreev reflections, respectively, are given. The features of CAR and single-particle transport are compared in Sec.~\ref{secndndn}. Finally, conclusions are drawn in Sec.~\ref{conclusions}.

\section{Model}
\label{model}

The systems we consider are generally described by a Hamiltonian for a hybrid system composed by multiple quantum dots tunnel coupled to both normal and superconducting leads. 
Each individual dot, labeled by the index $r$,  is described by the Anderson-impurity model with an onsite interaction $U_\m{intra}$ and the  level energy $\e_{r\sigma}$.  The interaction between electrons in different dots is characterized by the inter-dot repulsion $U_\m{inter}$. 
The quantum dots are described by the Hamiltonian
\begin{align}\nonumber
H_\m{dot}=\sum \limits_{r \sigma}\e_{r\sigma}(t) n_{r \sigma}+U_\m{intra}\sum\limits_r n_{r \uparrow} n_{r \downarrow} \\ +
\frac{1}{2}U_\m{inter}\sum\limits_{r\ne r' \sigma \sigma'}n_{r\sigma}n_{r'\sigma'}\, ,
\end{align}
where  $n_{r \sigma}= d_{r \sigma}^\dagger d_{r \sigma}$ is the number operator for electrons in the dot $r$ with spin $\sigma$ and  $d_{r \sigma}$ ($d_{r \sigma}^\dagger$) is the corresponding annihilation (creation) operator.  
Here we explicitly introduce the time-dependence of the dot levels, which is used to realize the pumping cycles.
The leads are described by the Hamiltonian  
\begin{equation}
\label{leadham}
H_\m{\eta}=\sum\limits_{k\sigma}\e_{\eta k}c^\dagger_{\eta k\sigma}c_{\eta k\sigma}-\delta_{\eta S}\sum\limits_k\left(\Delta c_{\eta-k\downarrow}c_{\eta k\uparrow}+\m{h.c.}\right)
\end{equation}
where the different reservoirs are identified by the label $\eta$. The operator $c_{\eta k\sigma}$ ($c^\dagger_{\eta k\sigma}$) annihilates (creates) an electron with momentum $k$ and spin $\sigma$ in lead $\eta$. The second term is only present for the superconducting leads and it is simply the attractive potential of  the mean-field BCS Hamiltonian. Without loss of generality the pair potential $\Delta$ can be chosen to be real, because we consider only one superconductor. 
Finally, the dots are coupled to the different leads by means of the tunneling Hamiltonian 
\begin{equation}
H_\m{tunn}=\sum\limits_{\eta r k \sigma}t_{\eta r} c^\dagger_{\eta k \sigma}d_{r\sigma} +\m{h.c.}\, .
\end{equation}
Both the tunnel matrix elements and the density of states of the leads $\rho_\eta$ are chosen to be energy independent in the window relevant for transport. Tunnel-coupling strengths are then defined as $\Gamma_{\eta,r,\sigma}=2\pi|t_{\eta,r}|^2 \rho_{\eta,\sigma}$. Notice that no inter-dot tunneling is included in the model.
Finally, the total Hamiltonian for this type of hybrid system  can be written as $H=H_\m{dot}+H_\m{tunn}+ \sum_\eta H_\m{\eta}$. We set in the following $\hbar=1$.

\subsection{Double-dot device}
The main focus of this paper is on the parallel double-dot device shown in Fig.~\ref{setup}, that is ideal for studying Cooper-pair splitting. It is composed of two quantum dots which are tunnel-coupled to different normal conductors but the same superconducting lead. We will refer to it as to the NDSDN system, where N indicates a normal lead, S a superconducting lead and D a quantum dot. 
The Hamiltonian of the NDSDN system is obtained from the general Hamiltonian of the previous subsection by having  $r\in\{L,R\}$,  $\eta\in N_L,N_R, S$ and $\Gamma_{N_L}\equiv \Gamma_{N_L, L}$, $\Gamma_{N_R}\equiv \Gamma_{N_R, R}$, $\Gamma_{S, r}$ as spin-independent tunnel-coupling strengths. With this we define $\Gamma_N\equiv\Gamma_{N_L}+\Gamma_{N_R}$. For the double-dot system we assume the dots' levels to be spin degenerate, that is $\e_{r\uparrow}=\e_{r\downarrow}=\e_r$, 
the Coulomb repulsion within one dot to be infinite $U_\m{intra}\rightarrow\infty$, and a finite inter-dot interaction $U_\m{inter}\equiv U$.
The limit $U_\m{intra}\rightarrow\infty$ excludes the possibility of double occupation of the same dot and, therefore, only CAR and no local AR appears.
As independent pumping parameters we choose the two spin-degenerate dot levels, $\{\e_{L},\e_R\}$, which can be varied by means of gate voltages. This system will be contrasted to the system with the lead S in its normal state, which is referred to as NDNDN and in which we take $\eta\in N_L,N_R, N_c$. 

\subsection{Single-dot device}
In order to identify the processes relevant for pumping, we consider a single-level quantum dot tunnel coupled to a ferromagnet and a superconductor (FDS), which having a smaller Hilbert space allows for a simpler analysis. 
The Hamiltonian  of the single-dot system is obtained from the general Hamiltonian considering only one dot (we consistently drop the index $r$) and two leads: $\eta\in F, S$. The ferromagnet is described by the Stoner model which induces $\Gamma_\uparrow\ne\Gamma_\downarrow$. The nonvanishing tunnel-coupling strengths are: $\Gamma_F$ and $\Gamma_S$. The pumping cycle in this case is realized by varying independently the two spin-split levels $\e_\uparrow, \e_\downarrow$. This can be done by means of a time-dependent gate voltage and magnetic field.  

\subsection{Large-$\Delta$ limit}
\label{sec_eff_H}
In the $\Delta\rightarrow\infty$ limit quasi-particle transport in the superconducting lead is suppressed and an effective description of the dot that takes into account Andreev tunneling can be obtained by 
integrating out the superconducting degrees of freedom.~\cite{effham,futterer10,eldridge10} Here we will discuss the resulting effective Hamiltonian only for the NDSDN system. The one for the FDS system is completely analogous. 
The effective Hamiltonian in the limit $U_\m{intra}\rightarrow\infty$ reads~\cite{eldridge10}
\begin{align}
H_\m{eff}&=\sum\limits_{r \sigma}\e_r n_{r \sigma}+U\sum\limits_{\sigma \sigma'}n_{L\sigma}n_{R\sigma'}\nonumber\\
 &+\frac{1}{2}\Gamma_{S}\left(d^\dagger_{R\uparrow}d^\dagger_{L\downarrow}-d^\dagger_{R\downarrow}   
d^\dagger_{L\uparrow}+\m{h.c.}\right)
   \end{align}
with $\Gamma_S=\sqrt{\Gamma_{SL}\Gamma_{SR}}$ being the effective coupling. The eigenstates are $\ket{\chi}\in\{\ket{+},\ket{-},\ket{\sigma,0},\ket{0,\sigma},\ket{T_{-1}},\ket{T_0},\ket{T_1}\}$ , where $\ket{\sigma,0}$ ($\ket{0,\sigma}$) corresponds to the left (right) dot being singly occupied with spin $\sigma$ and the right (left) dot being empty. The triplet states are  $\ket{T_{-1}}=\ket{\downarrow,\downarrow}$,  $\ket{T_0}=\left(\ket{\downarrow,\uparrow}+\ket{\uparrow,\downarrow}\right)/\sqrt{2}$ and $\ket{T_{1}}=\ket{\uparrow,\uparrow}$. The tunnel-coupling to the superconductor leads to eigenstates that are coherent superpositions of the state with both dots empty $\ket{0,0}$ and the singlet state $\ket{S}=\left(\ket{\downarrow,\uparrow}-\ket{\uparrow,\downarrow}\right)/\sqrt{2}$,
\begin{align}
\ket{\pm}=\frac{1}{\sqrt{2}}\sqrt{1\mp\frac{\delta}{2\e_A}}\ket{0}\mp\frac{1}{\sqrt{2}}\sqrt{1\pm\frac{\delta}{2\e_A}}\ket{S}\, ,
\end{align}
where $\delta\equiv\e_L+\e_R+U$  is the detuning between the empty state and the singlet and  $2\e_A\equiv\sqrt{\delta^2+2\Gamma_S^2}$ is the energy splitting between the $\ket{+}$ and $\ket{-}$ states.
 The corresponding eigenenergies are $E_{\pm}=\frac{\delta}{2}\pm\e_A$, $E_{(\sigma,0)}=\e_L$, $E_{(0,\sigma)}=\e_R$, and $E_{T_{-1}}=E_{T_{0}}=E_{T_{1}}=\e_L+\e_R+U$. In the FDS setup the eigenenergies and eigenstates are the same except that $L$ and $R$ are replaced by $\uparrow$ and $\downarrow$, respectively, the triplet state does not exist, the singlet state $\ket{S}$ is replaced by a double occupation $\ket{d}=d^\dagger_\uparrow d^\dagger_\downarrow\ket{0}$ of the dot, and $2\e_A\equiv\sqrt{\delta^2+\Gamma_S^2}$.

\section{Method}
\label{method}

\subsection{Generalized master equation}

The system, which is described by the Hamiltonian given in the previous section, can be subdivided into two different subsystems, the (proximized) quantum dots and the normal conducting leads. Since we are not interested in the dynamics of the leads' degrees of freedom, we can trace them out. This leads to an effective description of the quantum dots in terms of the reduced density matrix $\rho_\m{red}$. The elements of this reduced density matrix are denoted by  $p^{\chi_1}_{\chi_2}=\bra{\chi_1}\rho_\m{red}\ket{\chi_2}$, where $\chi_1$ and $\chi_2$ are states of the dots. The diagonal elements $p_\chi\equiv p^\chi_\chi$ give the probability to find the dots in state $\chi$. We introduce the vector $\boldsymbol{\pi}=(p_{\chi_1},...,p_{\chi_m},...,p_{\chi_j}^{\chi_i},...)^\m{T}$, with $i\neq j$, where the first $m$ components are the diagonal elements of the reduced density matrix of an $m$-dimensional Hilbert space followed by the off-diagonal elements. 
The dynamics of the reduced density matrix is governed by a generalized master equation (in matrix notation)
\begin{equation}\label{master}
	\frac{d}{d t}{\boldsymbol\pi}(t)=-\m{i}{\bf E}(t){\boldsymbol\pi}(t)+\int\limits^t_{-\infty}d t'{\bf W}(t,t'){\boldsymbol\pi}(t')\, .
\end{equation}
The matrix elements $W_{\chi'\chi'''}^{\chi\,\,\,\chi''}(t,t')$ of the kernel describe transitions  from an initial state at time $t'$ described by $p_{\chi'''}^{\chi''}$ to a final state at time $t$ described by $p_{\chi'}^{\chi}$.  In the systems which we consider, consisting of a single dot or two dots coupled in parallel, the matrix elements of ${\bf E}(t)$ are given by $E_{\chi'\chi'''}^{\chi\chi''}(t)=\delta_{\chi\chi''}\delta_{\chi'\chi'''}\left(E_\chi(t)-E_{\chi'}(t)\right)$. 

We study transport relying on the periodic variation of a set of  pumping parameters $\{X_i(t)\}$. Assuming the parameter modulation to be slow, that is the pumping frequency $\Omega$ to be small compared to all other energies of the system, we can perform an adiabatic expansion of Eq.~(\ref{master}) following the lines of Ref.~\onlinecite{splettstoesser06}.  Within the adiabatic expansion with respect to a reference time $t$, the reduced density matrix is written as the sum of an instantaneous contribution and its adiabatic correction, ${\boldsymbol \pi}(t)\rightarrow {\boldsymbol \pi}_t^{(i)}+{\boldsymbol \pi}_t^{(a)}$. The instantaneous contribution results from freezing all parameters to their value at time $t$ and yields a contribution in zeroth order in $\Omega/\Gamma_N$, indicated by the superscript $(i)$. The fact that the actual state of the system always slightly lags behind the parameter modulation is captured in the adiabatic term of first order in $\Omega/\Gamma_N$, indicated by $(a)$.\\
On top of the adiabatic expansion we perform a systematic expansion in the weak tunnel-coupling strengths between normal conductor and leads,  $\Gamma_{N}< k_BT$, of the kernel and the reduced density matrix, taking into account tunneling processes up to first order in $\Gamma_N$.  Orders in the perturbation expansion in the tunneling coupling are denoted by numbers in the superscript.
The instantaneous contribution to the reduced density matrix is determined by
\begin{equation}\label{eq_p_inst}
0=\left( -\m{i}{\bf E}(t)+{\bf W}_t^{(i,1)}\right){\boldsymbol \pi}_t^{(i,0)}
\end{equation}
together with the normalization condition $\boldsymbol n \boldsymbol \pi_t^\m{(i,0)}=1$ with $\boldsymbol n=(1,...1,0,...,0)$, that is, the first $m$ components of $\boldsymbol n$ are $1$ and the other components are $0$. The Laplace transform of the Kernel at zero frequency, with all parameters frozen to the time $t$ is given by ${\bf W}_t^{(i)}\equiv\lim\limits_{z\rightarrow 0^+}\int_{-\infty}^t dt'e^{-z(t-t')}{\bf W}_t^{(i)}(t-t')$, where here we consider only the first order in $\Gamma_N$ (if not specified otherwise). The adiabatic correction to the reduced density matrix turns out to have a contribution in minus first order in $\Gamma_N$,~\cite{splettstoesser06} which due to the adiabaticity condition $\Omega/\Gamma_N\ll1$ is not divergent and it is given by
\begin{equation}\label{eq_p_ad}
\frac{d}{dt}{\boldsymbol \pi}_t^{(i,0)}=\left( -\m{i}{\bf E}(t)+{\bf W}_t^{(i,1)}\right){\boldsymbol \pi}_t^{(a,-1)}
\end{equation}
with ${\boldsymbol n}{\boldsymbol \pi}^{(a,-1)}=0$. The rates ${\bf W}_t^{(i,1)}$ between diagonal elements of the reduced density matrix can be obtained by means of Fermi's Golden Rule. Solely for the ones connecting off-diagonal elements this is not sufficient and one has to resort to a diagrammatic method which has been developed in Refs.~\onlinecite{konig96}. In general, offdiagonal elements of the reduced density matrix, $p_{\chi'}^{\chi}$, enter Eqs.~(\ref{eq_p_inst}) and (\ref{eq_p_ad}). However, we assume weak coupling to the normal conductors $\Gamma_{N}\ll k_B T,\e_A$, where for the FDS as well as the NDSDN setup the offdiagonal elements of the reduced density matrix are decoupled from the diagonal ones.~\cite{leijnse08}  As we are interested in the diagonal elements, needed for the computation of the current, we can therefore disregard the offdiagonal ones. Solely in the NDNDN setup the dynamics of the offdiagonal elements $p^{(\sigma,0)}_{(0,\sigma)}$ and $p_{(\sigma,0)}^{(0,\sigma)}$ is coupled with the dynamics of the occupation probabilities.
In the NDNDN system, where also offdiagonal elements of the reduced density matrix contribute, we assume $\Delta\e=\e_L-\e_R\approx\Gamma_N$ and ${\bf E}$ and ${\bf W}_t^{(i,1)}$ have to be of the same order in the small parameter $\Gamma_N\simeq\Delta\e$.~\cite{wunsch05,riwar10} 

In a similar way, one can write rate equations for the expectation value of the current into lead $\eta$. The instantaneous contribution to the current  is
\begin{equation}\label{eq_I_inst}
I^{(i)}_\eta(t)=e{\bf n}{\bf W}_t^{\eta,(i)}{\boldsymbol \pi}^{(i)}_t \, ,
\end{equation}
which we consider in first order in the tunnel coupling, only. The current rates ${\bf W}^\eta_t$ take into account  the number of electrons transferred to lead $\eta$.~\cite{konig96} From Eq.~(\ref{eq_I_inst}), we derive the conductance, which is given by $G=(dI^{(i,1)}/dV)|_{V=0}$, with $V$ being the bias voltage. The instantaneous current vanishes exactly in the absence of an applied bias.
The adiabatic correction to the current is then the dominant one and it is given by
\begin{equation}
\label{adiabstrom}
I^{(a,0)}_\eta(t)=e{\bf n}{\bf W}_t^{\eta,(i,1)}{\boldsymbol \pi}^{(a,-1)}_t\, .
\end{equation}
We are in the following interested in the charge transferred into lead $\eta$ per cycle of the parameter variation. This is found by integrating the current over one period
\begin{equation}
Q_{X_1,X_2}^{\eta}=\int_0^{2\pi/\Omega} dt I^{(a,0)}_\eta(t).
\end{equation}

In the following we consequently drop the index $\eta$ if the pumped charge corresponds to the superconductor, $Q_{X_1,X_2}\equiv Q_{X_1,X_2}^S$. Two time-dependent parameters are necessary to create a nonvanishing pumped charge; we indicate the parameter choice in the subscript. The pumping parameters can be written as $X_i(t)=\overline{X}_i+\delta X_i(t)$, where $\overline{X}_i$ is the mean value and $\delta X_i(t)$ the oscillating component. We concentrate on the limit of weak pumping, that is, the oscillating component is small compared to the tunnel coupling $\delta X_i(t)\ll \Gamma_N$. Therefore, we only account for terms up to bilinear order in $\delta X_i(t)$ and the pumped charge is proportional to $A_{X_1,X_2}=\int_0^{2\pi/\Omega}dt\delta X_1(t)\frac{d}{dt}\delta X_2(t)$.

\section{results}

Using the effective Hamiltonian and performing  the perturbation expansion as presented in the previous section we calculate the pumped charge in lowest order in $\Gamma_N$ or $\Gamma_F$, respectively. Close to the dot levels being at resonance, the lowest order processes are the dominant ones and cotunneling processes can safely be neglected.  Before tackling the more complicated problem of CAR, we will  first study the FDS system, in order to understand the features of local AR in adiabatic pumping and to identify the different transport processes occurring in this simple setup. For this setup, we also examine the influence of cotunneling processes on the pumped charge far from resonance (Coulomb-blockade regime), which are important when the interaction $U$ becomes much larger than the temperature. 
In Sec.~\ref{secndsdn}, we discuss how adiabatic pumping provides the possibility to study CAR. To this end, we finally compare the NDSDN setup with the NDNDN setup.

\subsection{Local Andreev reflection}
\label{secnds}
In this section we consider adiabatic pumping through the FDS setup. We choose the dot-level positions for electrons with different spins  $\e_\uparrow(t)$ and $\e_\downarrow(t)$ to be the pumping parameters. Such a situation can be realized by a time-dependent gate voltage and a time-dependent magnetic field, the latter introducing a time-dependent Zeeman splitting. This choice of pumping parameters is convenient here as it allows for a direct comparison with a double dot in the absence of a magnetic field, in which gate voltages applied to the two dots are independently modulated. Pumping is possible whenever the polarization of the leads or the average level splitting  $\overline{\Delta\e}\equiv\overline{\e}_\uparrow-\overline{\e}_\downarrow$ are nonvanishing. To get a better understanding of the transport properties we first focus on two different limits: a vanishing polarization ($p=0$) and a vanishing average level splitting ($\overline{\e}_\uparrow=\overline{\e}_\downarrow$). We start with the case of a vanishing polarization and a finite level splitting.

For the pumped charge we find
\begin{widetext}
\begin{align}
\label{Qfdsp0}
Q_{\e_\uparrow,\e_\downarrow}(p=0)\approx-\frac{eA_{\e_\uparrow,\e_\downarrow}\Gamma_S^2}{\left[\Gamma_S^2+(U+\overline{\e}_\uparrow+\overline{\e}_\downarrow)^2\right]^\frac{3}{2}}
\frac{f(\overline{E}_--\overline{\e}_\uparrow)f'(\overline{E}_--\overline{\e}_\downarrow)-f(\overline{E}_--\overline{\e}_\downarrow)f'(\overline{E}_--\overline{\e}_\uparrow)}{\left[f(\overline{E}_--\overline{\e}_\uparrow)+f(\overline{E}_--\overline{\e}_\downarrow)-f(\overline{E}_--\overline{\e}_\uparrow)f(\overline{E}_--\overline{\e}_\downarrow)\right]^2}
\end{align}
\end{widetext}
with $f'(x)=\frac{d}{dx}f(x)$ being the derivative of the Fermi function. We made use of the approximation $f({E}_+-{\e}_\uparrow)\approx f({E}_+-{\e}_\downarrow)\approx 0$ and $f({\e}_\uparrow-{E}_+)\approx f({\e}_\downarrow-{E}_+)\approx 1$, which is justified for $\Gamma_S > k_BT$. Equation~(\ref{Qfdsp0}) shows that the pumped charge vanishes for an average Zeeman splitting equal to zero, that is $\overline{\e}_\uparrow=\overline{\e}_\downarrow$. 

\begin{figure}
\includegraphics[width=.45\textwidth]{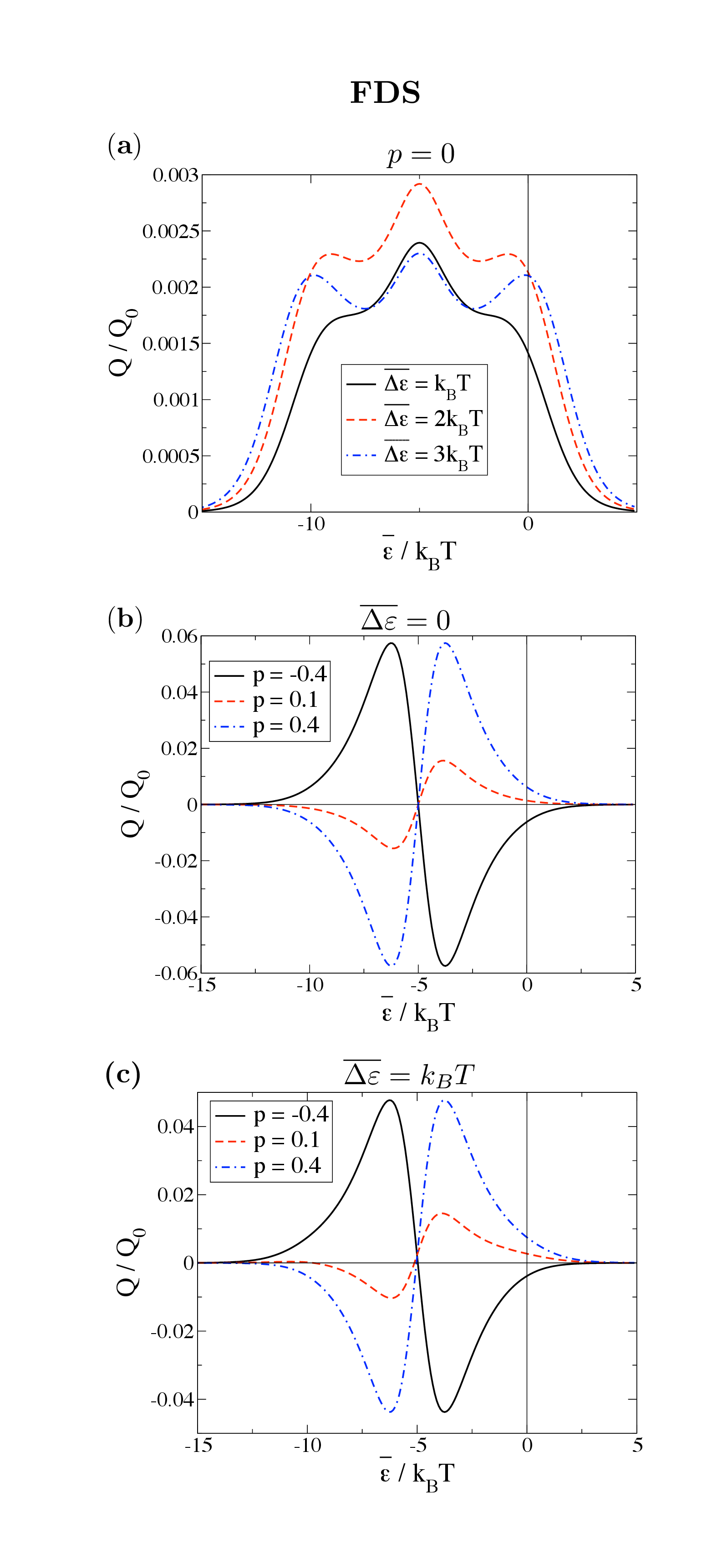}
\caption{
\label{figfds}
(Color online)  Pumped charge $Q\equiv Q_{\e_\uparrow,\e_\downarrow}$ in units of $Q_0=\frac{eA_{\e_\uparrow,\e_\downarrow}}{\left(k_BT\right)^2}$ as a function of the average dot level $\overline{\e}$. The parameters in all figures are $\Gamma_S=4k_BT$ and $U=10k_BT$.}
\end{figure}

In Fig.~\ref{figfds}(a), we show 
the pumped charge $Q_{\e_\uparrow\e_\downarrow}$, without  the approximation on the Fermi functions used to write Eq.~(\ref{Qfdsp0}), as function of the average value of the mean dot level $\overline{\e}\equiv(\overline{\e}_\uparrow+\overline{\e}_\downarrow)/2$.  
The pumped charge exhibits a three-peak structure. The two external peaks are observed when the dot is in resonance with the normal lead, that is when the addition energy for a single electron equals the chemical potential. This is realized for ${E}_{-}-{\e}_\sigma=0$.  Since we consider Zeeman splitting $\Delta\e$  being larger than $k_B T$ (with ${\e}_\downarrow$ being the level with the lower energy), only the resonance ${E}_{-}-{\e}_\downarrow=0$ is accessible due to Coulomb blockade. The other Andreev bound state, with energy $E_+$, is only accessible in the high-bias or high temperature regime. 
The two resonances associated to the condition ${E}_{-}-{\e}_\downarrow=0$  are at the two positions, $\e_\m{max,1\pm}\approx \frac{1}{2}(-U\pm[\left(U+{|\Delta\e|}\right)^2-\Gamma_S^2]^{1/2})$, and are enhanced for an increased average Zeeman splitting.  

The central peak appears when the dot is in resonance with the  superconductor, that is the average dot level is $\overline{\e}_\m{max,2}\approx -U/2$ which is realized for $\delta<\Gamma_N$. In this case the dot undergoes fast oscillations between the empty and doubly-occupied state due to coherent Cooper-pair transfer. In particular these oscillations are much faster than tunneling events of single particles between the  normal conductor and the dot. 
However, transport requires exchange of charge both with the normal and the superconducting leads. Therefore, increasing the Coulomb repulsion $U$ leads to an overall suppression of the pumped charge. The three peaks are not suppressed in the same manner. The side peaks are suppressed by the factor $\left[1+(U+\overline{\e}_\uparrow+\overline{\e}_\downarrow)^2/ \Gamma_S^2\right]^{-\frac{3}{2}}$, appearing in Eq.~(\ref{Qfdsp0}). Instead the central peak is suppressed by the combination of Fermi functions in Eq.~(\ref{Qfdsp0}).

We now focus on the limit of a vanishing average level splitting ($\overline{\Delta\e}=0$) and a finite polarization. The pumped charge is then given by
\begin{align}
\label{Qfdsde0}
Q_{\e_\uparrow,\e_\downarrow}(\overline{\Delta\e}=0)\approx\frac{4eA_{\e_\uparrow,\e_\downarrow}p\left(1-p^2\right)\Gamma_S^2\delta}{k_BT\left[\Gamma_S^2+(1-p^2)\delta^2\right]^2}\cdot \frac{1-f(\overline{E}_--\overline{\e})}{2-f(\overline{E}_--\overline{\e})}
\end{align}
approximating the Fermi functions as done above. We find that the pumped charge is an odd function of $\delta$, therefore vanishing at the electron-hole symmetric point.
 The full result for the pumped charge at zero average detuning, $\overline{\Delta\e}=0$, is shown in Fig.~\ref{figfds}(b). As shown in Eq.~(\ref{Qfdsde0}), the pumped charge vanishes at $\overline{\e}=-U/2$. However, we find a peak-trough structure, that is, the maximum contribution to the pumped charge appears in two peaks, close to $\overline{\e}\approx-U/2$, with opposite sign. As argued above this relies on fast Cooper-pair oscillation. The amplitude of the pumped charge is much larger than in Fig.~\ref{figfds}(a) and strongly depends on the polarization of the leads: the stronger the polarization the larger the amplitude. Furthermore, the pumped charge, in the vicinity of the electron-hole symmetric point is not  suppressed by the strong Coulomb repulsion. We will address this, when discussing the cotunneling regime.

Instead of giving the lengthy expression of the pumped charge for a finite average level splitting, $\overline{\Delta\e}\neq0$, and a finite polarization, $p\neq0$, we show it in Fig.~\ref{figfds}(c) as a function of the average dot level. The shape is a combination of the two structures shown in Figs.~\ref{figfds}(a) and \ref{figfds}(b). We find that  the effect for the finite polarization dominates. Therefore, the peaks around $\overline{\e}\approx-U/2$ with opposite sign are the main feature to identify the proximization of the dot. 

When lowering the temperature, the height of the peak-trough structure increases with inverse temperature, that is, it becomes more and more pronounced.
This result can, however, only be trusted as long as temperature is still large enough such that all charge states are thermally occupied. 
In the Coulomb-blockade regime, $U\gg k_\m{B}T$ and $\delta\equiv\e_\uparrow+\e_\downarrow+U\approx k_\m{B}T$, when the sequential tunneling rates to reach an empty or doubly-occupied dot are exponentially small, higher-order processes such as cotunneling need to be taken into account.
To compare with the result presented in Eq.~(\ref{Qfdsde0}), we analyze the pumped charge in the cotunneling regime. 
For this, we first of all note that Eq.~(\ref{eq_p_ad}) looses its validity in the Coulomb-blockade regime, since the rates $W^{(i,1)}$ get exponentially suppressed, while  - in contrast to situations where the magnetic field is constant~\cite{splettstoesser06} - the time-derivative of the instantaneous occupation probabilities, $\frac{d}{dt}p_{t,\sigma}^{(i,0)}$, of single occupation with spin $\sigma$ do not. The time-evolution of the probabilities of single occupation is then governed by 
spin-flip processes in second order in the tunneling, $W_{t,\downarrow\uparrow}^{(i,2)}$, entering Eqs.~(\ref{eq_p_ad}) and (\ref{adiabstrom}) together with adiabatic corrections to the probability in minus second order in $\Gamma$, $p^{(a,-2)}_{t,\sigma}$. However, since $U\gg k_\m{B}T$ and $\delta\approx k_\m{B}T$ results in an exponential suppression of $\frac{d}{dt}p_{t,\pm}^{(i,0)}$, also the elements $p^{(a,-2)}_{t,\pm}$ are suppressed and will not enter the current in the Coulomb blockade regime.
For the calculation of the cotunneling rates we follow the procedure introduced in Refs.~\onlinecite{averin89} for metallic islands and applied for single-level quantum dots, for example, in Ref.~\onlinecite{weymann05/1}. In contrast to Eq.~(\ref{adiabstrom}) in the cotunneling regime the current is then 
\begin{equation}\label{eq_cot}
I^{(a,0)}_F(t)=e \left[W_{t,\downarrow\uparrow}^{F,(i,2)}p^{(a,-2)}_{t,\uparrow}+W_{t,\uparrow\downarrow}^{F,(i,2)} p^{(a,-2)}_{t,\downarrow}\right]\, ,
\end{equation}
which is nonvanishing due to $p^{(a,-2)}_{t,\uparrow}=-p^{(a,-2)}_{t,\downarrow}$ and $W_{t,\downarrow\uparrow}^{F,(i,2)}=-W_{t,\uparrow\downarrow}^{F,(i,2)}$. Due to charge conservation, $Q_{\e_\uparrow,\e_\downarrow}=-Q_{\e_\uparrow,\e_\downarrow}^F$, the charge pumped into the superconductor is found as
\begin{align}
&Q_{\e_\uparrow,\e_\downarrow}\approx\nonumber\\ 
&-\frac{3e\pi^2k_BT A_{\e_\uparrow,\e_\downarrow}\Gamma_S^2p\left(1-p^2\right)\delta\left(\Gamma_S^2+\delta^2-U^2\right)}{\left[(1+p^2)\pi^2(k_BT)^2\Gamma_S^2+\frac{3}{32}(1-p^2)\left(\Gamma_S^2+\delta^2-U^2\right)^2\right]^2}\, , 
\end{align}
where we used $\overline{\Delta\e}/U\ll1$.
The qualitative behavior of the pumped charge in the cotunneling regime strongly differs from the sequential tunneling regime. For strong Coulomb interaction, in the cotunneling regime transport is suppressed with $1/U^6$. To find a possible explanation for this suppression we focus on the transport processes during one pumping cycle. Consider the following process where a net transport is obtained in the cotunneling as well as in the sequential tunneling regimes: An electron tunnels from the ferromagnet onto a singly occupied dot. The dot is then, for example, in state $\ket{-}$. To obtain a net transport another electron has to tunnel from the ferromagnet onto the quantum dot bringing it back into single occupation which is possible due to Cooper-pair oscillations.  A comparison of the system's time scales for the two regimes 
might shed light on the origin of the suppression of the pumped charge. In the sequential tunneling regime the time between two single-electron transport processes scales with $1/\Gamma_N$. In the cotunneling regime the intermediate state can only be virtually occupied due to energy conservation and hence the time between two tunneling events scales with  $1/U$. In the considered limit of  large $U\gg k_BT$ and small $\Gamma_N\ll k_BT$, Cooper-pair oscillations are fast compared to the time between two tunneling events in the sequential but slow in the cotunneling regime. This gives an interpretation of the suppression of the pumped charge in the cotunneling regime.

\subsection{Crossed Andreev reflection}
\label{secndsdn}
We now consider a system made out of two quantum dots each coupled to  one normal conducting lead. The two QDs are then coupled to each other via a common superconducting lead (see Fig.~\ref{setup}).  We take the pair potential in the superconducting lead to be the largest energy scale ($\Delta\rightarrow\infty$), such that single-particle transport between superconductor and QDs is suppressed. Furthermore, we take the intra-dot Coulomb repulsion ($U_\m{intra}\rightarrow\infty$) to be large excluding double occupation of each of the single dots, as discussed in Sec.~\ref{sec_eff_H}. In this regime, \textit{only nonlocal} effects enable transport between the superconductor and the dots, that is a Cooper-pair has to be split into two electrons occupying different dots or electrons from different dots enter the superconductor to form a Cooper-pair.\\
We now calculate the charge pumped through the system due to the periodic modulation of the dot levels $\e_\mathrm{L}(t)$ and $\e_\mathrm{R}(t)$, which can be achieved by two time-dependent gate voltages. We are interested in the charge, $Q_{\e_L,\e_R}$, pumped into the superconducting lead, which due to charge conservation and to the fact that only CAR is allowed is twice the charge pumped out of each normal lead.

\begin{figure}
\includegraphics[width=.44\textwidth]{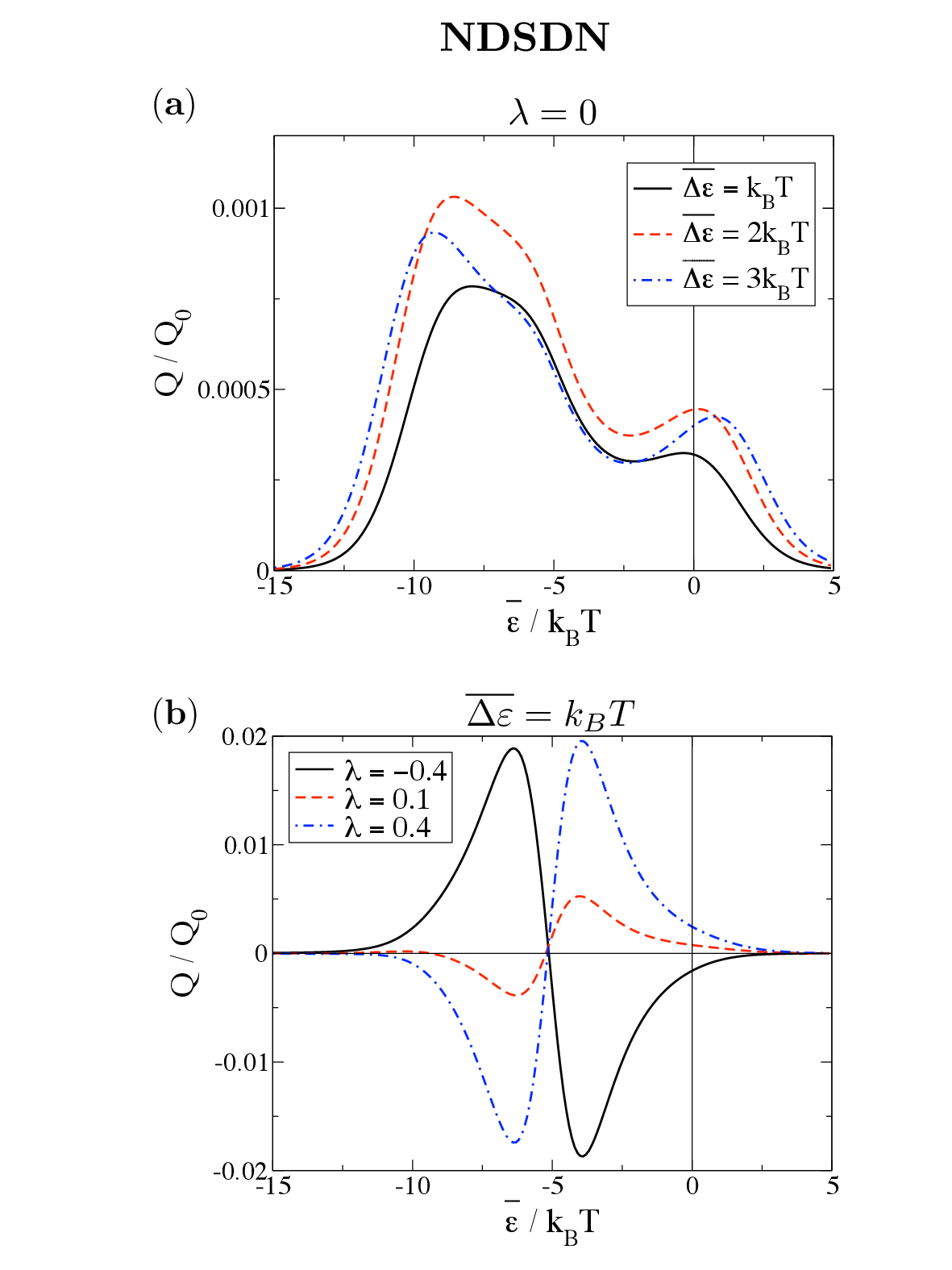}
\caption{
\label{figndsdnpump}
(Color online)  Pumped charge $Q\equiv Q_{\e_L,\e_R}$ in units of $Q_0=\frac{eA_{\e_L,\e_R}}{\left(k_BT\right)^2}$ as a function of the average dot level $\overline{\e}$. The parameters are $\Gamma_S=3k_BT$ and $U=10k_BT$.}
\end{figure}

 In Figs.~\ref{figndsdnpump}(a) and \ref{figndsdnpump}(b) we show $Q_{\e_L,\e_R}$  as a function of $\overline{\e}$  for different values of $\overline{\Delta\e}$ and for different coupling asymmetries with the normal conducting leads, $\lambda=(\Gamma_{N_L}-\Gamma_{N_R})/\Gamma_N$, respectively. 
\begin{figure}
\includegraphics[angle=270,width=.39\textwidth]{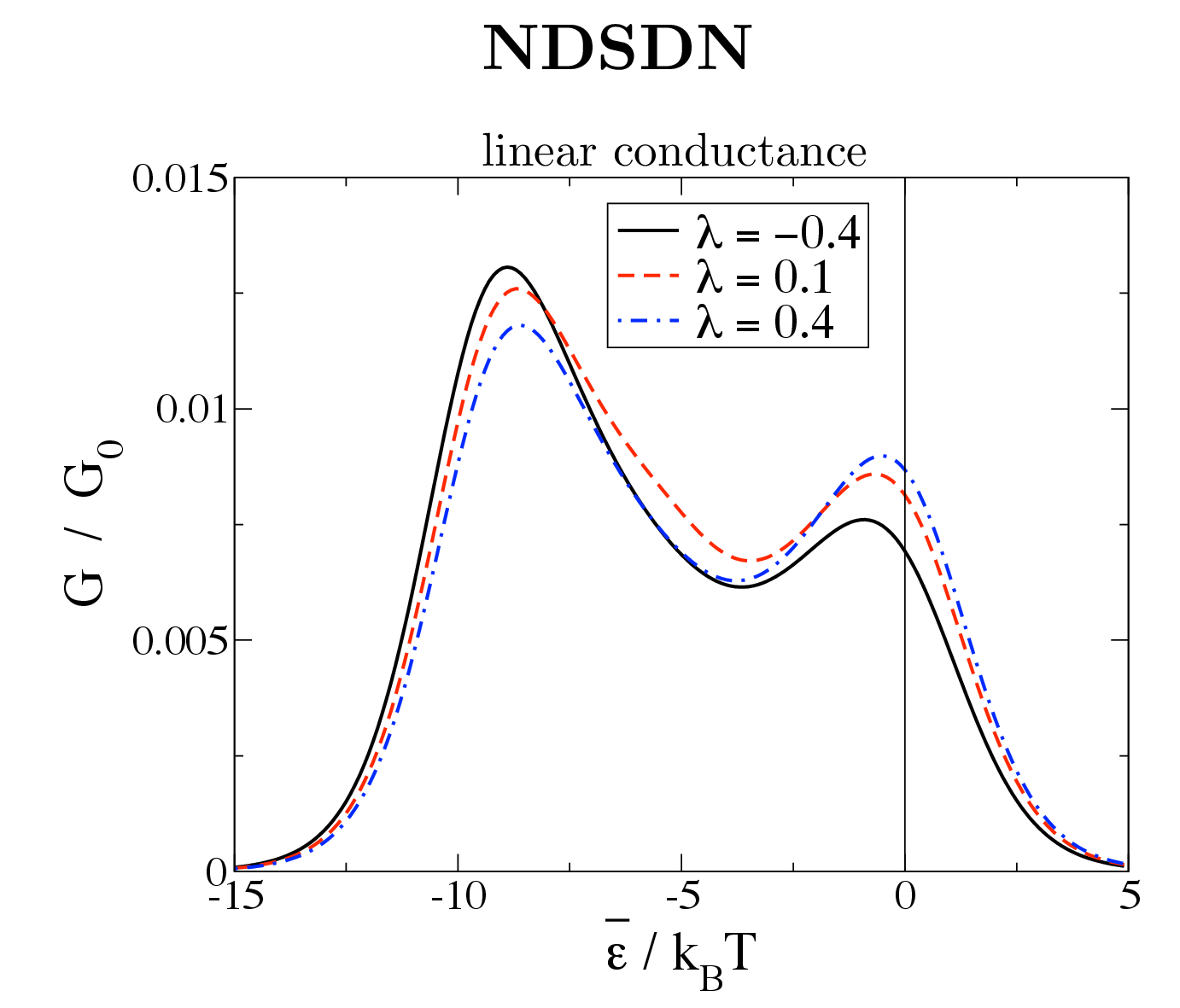}
\caption{
\label{ndsdn_cond}
(Color online) Linear conductance as a function of the average dot level $\overline{\e}$. for different coupling asymmetries $\lambda$. The parameters are $\Gamma_S=3k_BT$, $\Gamma_N=k_BT$, $U=10k_BT$, and $\overline{\Delta\e}=k_BT$.
}
\end{figure}
Features appear at the resonance condition with the normal and superconducting leads, that are equivalent to the one in the FDS case with Zeeman splitting replaced by the difference of the energy levels of the left and right dots and the  polarization $p$ replaced by coupling asymmetry $\lambda$. If the couplings to the normal leads are symmetric, $\lambda=0$, the charge as a function of the average mean dot level position $\overline{\e}$, shows three peaks similarly to the FDS case. In this respect, CAR exhibits similar features to AR through the single dot. The main difference between the two is the asymmetry in the heights of the external peaks which can be attributed to the triplet blockade discussed in Ref.~\onlinecite{eldridge10}. Since the proximization by the superconductor solely causes a coupling between the empty and the singlet state, Cooper-pair tunneling is blocked whenever the dot is in the triplet state. In the FDS setup the symmetry of the two external peaks can be related to particle-hole symmetry which is broken by this triplet blockade in the NDSDN structure.

As in the FDS with finite polarization, also in the NDSDN the scenario changes completely in the asymmetric-coupling case ($\lambda\neq 0$). In this case the peak at $\overline{\e}=-U/2$ is replaced by a large peak-trough structure. Interestingly, this feature dominates the external peaks which are barely visible in Fig.~\ref{figndsdnpump}(b). The position of the maxima and minima of this feature are exchanged when reversing the coupling asymmetry ($\lambda \rightarrow -\lambda$).

However, in the linear conductance, the coupling asymmetry does not introduce any new feature, as shown in Fig.~\ref{ndsdn_cond}, where for different coupling asymmetries only the weight of the three peaks is influenced and not their polarity. Furthermore, the central peak is strongly suppressed. That means that the characteristic features of CAR in adiabatic pumping are not present in the linear conductance. As we will see in the next section these features are fundamental to distinguish single-particle transport from CAR.


\subsection{Single-particle transport}
\label{secndndn}
A finite pumped charge can be obtained by varying in time  the properties of the two spatially-separated dots  exclusively by nonlocal correlations. CAR has such a nonlocal character. However, there may be other nonlocal effects that can produce a finite pumped charge and, thus, mask the signal from CAR. In order to distinguish CAR from other nonlocal transport processes, we investigate single-particle transport in a NDNDN setup, where the superconductor in the NDSDN setup is replaced by a normal conductor. While in the NDSDN setup the nonlocality arises from CAR, in the NDNDN setup pumping is possible due to the formation of a coherent superposition of  states with one electron either in the left or the right dot. This superposition is generated by the tunnel coupling to the common normal lead. In contrast to the NDSDN setup, the coherent superposition is strongly suppressed if the difference of the two dot levels is large compared to temperature ($|\Delta\e|\gg k_BT$).  

Furthermore, in the NDSDN setup pumping cannot lead to an average  charge  transfer  from  the left into the right normal lead (and vice versa) because transport through the superconductor always involves CAR in the infinite-$\Delta$ limit. Instead, in the NDNDN setup, charge can also be transferred from the left lead $N_L$ to the right lead $N_R$. Therefore an asymmetry of transport into lead $N_L$ and into lead $N_R$ is one possible indication for single-particle transport. 

\begin{figure}
\includegraphics[angle=270,width=.39\textwidth]{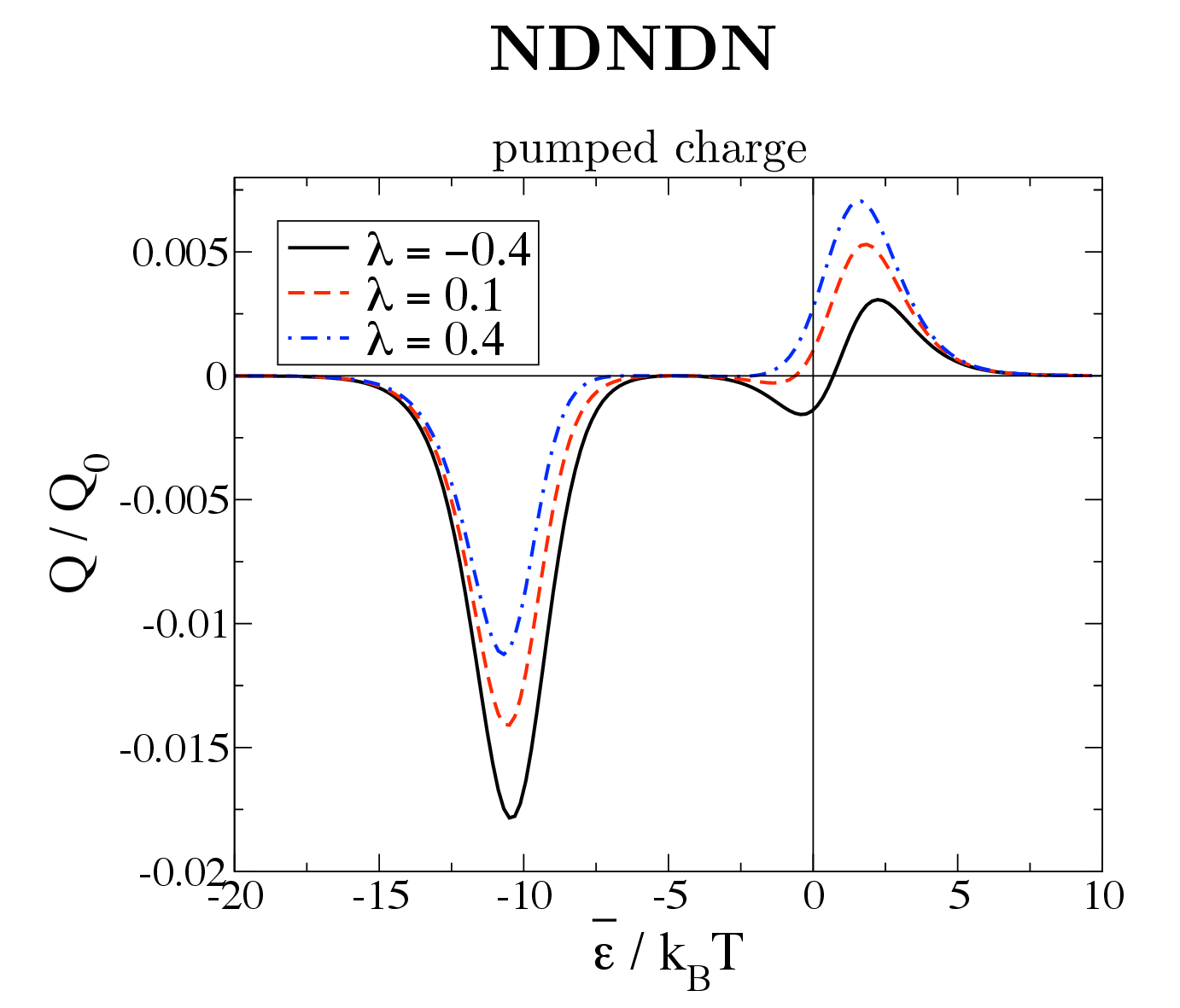}
\caption{
\label{ndndn}
(Color online) Pumped charge $Q\equiv Q^{N_c}_{\e_L,\e_R}$ in units of $Q_0=\frac{eA_{\e_L,\e_R}}{\left(k_BT\right)^2}$ as a function of the average dot level $\overline{\e}$ for  different coupling asymmetries $\lambda$. The other parameters are $\Gamma_{N_\m{c,L}}= 0.4 k_BT$,  $\Gamma_{N_\m{c,R}}=0.2k_BT$, $\Gamma_N=0.1k_BT$, $U=10k_BT$, and $\overline{\Delta\e}=k_BT$.}
\end{figure}

The motivation of this work is the identification of CAR with respect to quasi-particle transport in form of an easily detectable signature in the pumped charge. We find this to be the peak-trough structure at  $\overline{\e}=-U/2$  that appears in the presence of a coupling asymmetry. In  single-particle transport, modeled by the {NDNDN} setup, this feature is completely absent and only the peaks at the normal resonances appear (see Fig.~\ref{ndndn}). These normal resonances also have opposite signs, which cannot be reversed  by changing the coupling asymmetry. Also in the very special situation of a symmetric coupling ($\lambda=0$) CAR can be distinguished from single-particle transport by the presence of the peak at $\overline{\e}=-U/2$.
Therefore, an experimental study of the pumped charge in the double-dot system  as a function of $\lambda$ as well as its behavior around $\overline{\e}\approx -U/2$ can clearly distinguish CAR from quasiparticle transport.

\section{Conclusions}
\label{conclusions}
We have investigated adiabatic pumping through two quantum dots tunnel coupled to the same superconductor and additionally coupled to different normal conductors. For an infinite intra-dot Coulomb repulsion in this setup pumping relies on CAR. In order to understand the underlying transport processes we mapped the setup to the simpler setup of a quantum dot tunnel coupled to a ferromagnet and a superconductor where only AR appears. We found that most of the features of pumping including CAR are also present in pumping with local AR. The main difference are asymmetries due to the presence of the triplet state. To distinguish CAR from single-electron tunneling, which does not appear in our model but might be relevant in experiments, we compare transport through the double-dot setup containing a superconductor with a setup where the superconductor is replaced by a normal conductor. The dependence on the average dot-level position and the dependence on the coupling asymmetry $\lambda$ turn out to be the main distinguishing features.

\textit{Acknowledgments}. We acknowledge financial support from EU under Grant No. 238345 (GEOMDISS), the SFB 491, the SPP 1285, and KO 1987/5. J.S. acknowledges financial support from the Ministry of Innovation, NRW.

\end{document}